\documentclass[3p,times]{elsarticle}
\usepackage{lineno}
\usepackage{graphicx}
\usepackage{subfigure}
\usepackage{booktabs}
\usepackage[table]{xcolor}
\usepackage{multirow}
\usepackage{threeparttable}
\usepackage{amsmath}
\usepackage{amssymb}
\usepackage{engord}
\usepackage[colorlinks,linkcolor=blue]{hyperref}
\modulolinenumbers[5]

\journal{arXiv.org}

\begin{document}
\begin{frontmatter}

\title{Orderness Predicts Academic Performance: Behavioral Analysis on Campus Lifestyle}

\author[inst1]{Yi Cao}
\author[inst1]{Jian Gao}
\author[inst2]{Defu Lian}
\author[inst1]{Zhihai Rong}
\author[inst2]{Jiatu Shi}
\author[inst3]{Qing Wang}
\author[inst1]{Yifan Wu}
\author[inst2]{Huaxiu Yao}
\author[inst1,inst2]{Tao Zhou\corref{cor1}}
\cortext[cor1]{Corresponding author. \\ \hspace*{0.4cm} \emph{E-mail address}: \href{mailto:zhutou@ustc.edu}{zhutou@ustc.edu}}

\address[inst1]{CompleX Lab, Web Sciences Center, University of Electronic Science and Technology of China, Chengdu 611731, China}
\address[inst2]{Big Data Research Center, University of Electronic Science and Technology of China, Chengdu 611731, China}
\address[inst3]{Key Laboratory for NeuroInformation of Ministry of Education, School of Life Science and Technology, Center for Information in Medicine, University of Electronic Science and Technology of China, Chengdu 611731, China}

\begin{abstract}
Quantitative understanding of relationships between students' behavioral patterns and academic performances is a significant step towards personalized education. In contrast to previous studies that mainly based on questionnaire surveys, in this paper, we collect behavioral records from 18,960 undergraduate students' smart cards and propose a novel metric, called \emph{orderness}, which measures the regularity of campus daily life (e.g., meals and showers) of each student. Empirical analysis demonstrates that academic performance (GPA) is strongly correlated with orderness. Furthermore, we show that orderness is an important feature to predict academic performance, which remarkably improves the prediction accuracy even at the presence of students' diligence. Based on these analyses, education administrators could better guide students' campus lives and implement effective interventions in an early stage when necessary.
\end{abstract}

\end{frontmatter}

A major challenge in education management is to uncover underlying ingredients that affect students' academic performance, which is significant in working out teaching programs, facilitating personalized education, detecting harmful abnormal behaviors and intervening students' mentation, sentiments and behaviors when it is very necessary. For example, it has been demonstrated that physical status (e.g., height and weight) \cite{Jamison1986,Mosuwan1999,Taras2005,Stabler1994,Chang2002}, intelligence quotient (IQ) \cite{Deary2007,Laidra2007} and even DNA \cite{Krapohl2014,Okbay2016,Selzam2017} are correlated with educational achievement. Accordingly, we can design personalized teaching and caring programs for different individuals. Since we cannot change a student's height or DNA via education, more studies concentrate on the aspects of psychology and behavior, with a belief that learning problems resulted from psychological and behavioral issues can be at least partially intervened. For example, early interventions according to the predictions on course scores or course failures have been discussed recently for K12 education \cite{Bowers2013,Tamhane2014,Lakkaraju2015}.

Extensive experiments about relationships between personality and academic performance have been reported in the literature, suggesting that agreeableness, openness and conscientiousness, among the big five personality traits, are significantly correlated with tertiary academic performance, say GPA and course performance \cite{Connor2007,Poropat2009,Vedel2014}. In particular, the correlation between conscientiousness and GPA is the strongest \cite{Connor2007,Poropat2009,Vedel2014}. Behaviors are also associated with academic performance. The class attendance is known for long as an important determinant of academic performance \cite{Schmidt1983,Romer1993,Durden1995,Conard2006,Crede2010}, and the additional studying hours are positively correlated with GPA \cite{Crede2008,Stinebrickner2008,Grave2011}. In addition to the studying behaviors, some experimental evidences indicate that students with healthy lifestyles and good sleep habits have higher GPAs in average \cite{Trockel2000,Dewald2010,Taylor2013,Wald2014}. Under the traditional research framework, a large portion of data sets come from questionnaires and self-reports, which are usually of very small sizes (most sample sizes scale from dozens to hundreds, see meta-analysis reviews \cite{Poropat2009,Vedel2014,Crede2010,Crede2008,Dewald2010}) and suffer from social desirability bias \cite{Fisher1993,Paulhus2007}, resulting in the difficulties to draw valid and solid conclusions.

\begin{figure*}[htp]
	\centering
	\includegraphics[width=0.95\textwidth]{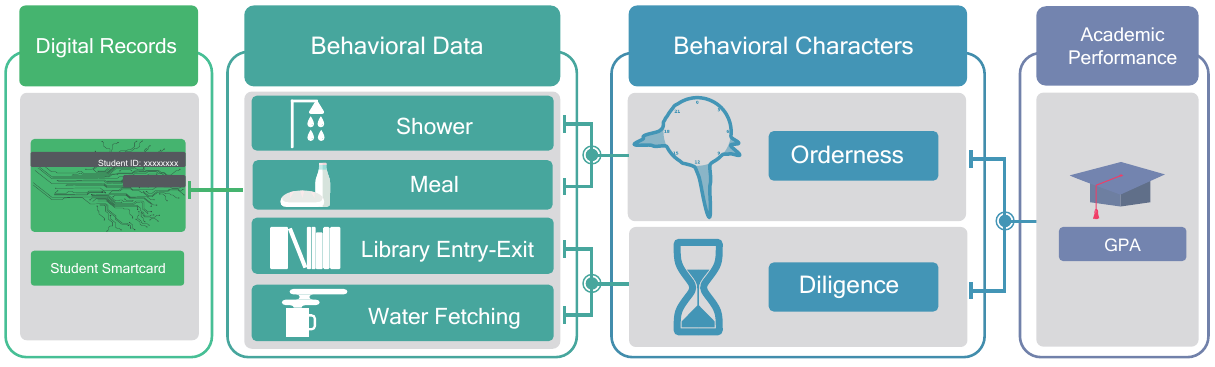}
	\caption{Methodology used to analyze correlations between campus daily routine and academic performance, and then to predict future academic performance. First of all, a large volume of digital entry-exit and consumption records are collected by the real-name campus smart cards with ID encryption. Then, four kinds of behaviors are used to measure two high-level behavioral characters: orderness and diligence. Specifically, taking showers in dormitories and having meals in cafeterias contribute to the orderness measure, while entering and exiting the library and fetching water in teaching buildings contribute to the diligence measure. After that, empirical analysis is performed to show the correlation between academic performance and behavioral characters (\emph{i.e.}, orderness and diligence). Last but not least, the predictive powers of orderness and diligence are also presented and compared.}
    \label{fig1}
\end{figure*}

Thanks to the fast development of modern information technology, we have unprecedented opportunities to collect real-time records of students' living and studying activities in an unobtrusive way, through smartphones \cite{Wang2014}, online courses \cite{Brinton2015}, campus WiFi \cite{Zhou2016}, and so on. Analyses on these data revealed many unreported correlations between behavioral features and academic performance. For example, playing more of the video than its length and pausing more than once are two strong indicators for better course performance in MOOCs \cite{Brinton2015}, and students who spend more time partying at fraternities or sororities have lower GPAs in average \cite{Wang2015}. In this paper, through campus smart cards, we have collected digital records of 18,960 undergraduate students' daily activities in the \emph{University of Electronic Science and Technology of China} (UESTC). We have extracted some high-level behavioral characters from the records, such as \emph{orderness} (evaluated by the purchase records for showers and meals) that quantifies the daily-life regularity and \emph{diligence} (evaluated by the entry-exit records in library and fetching water records in teaching buildings) that estimates how long time spent on studies. We have demonstrated significant correlations between orderness and GPA, as well as between diligence and GPA. In particular, orderness is for the first time to our knowledge proposed as an important behavioral character that largely affects a student's academic performance. Further analyses show that the introduction of orderness as a feature can remarkably improve the prediction accuracy of semester GPA at the presence of diligence. The methodology used in the present work is illustrated in Fig.~\ref{fig1}.

\section*{Results}
\subsection*{Orderness}
Intuitively, a regular lifestyle would stand us in good stead for college study. In particular, teachers and administrators in most Asian countries (\emph{e.g.}, Japan, Korea, Singapore, China, \emph{etc}.) ask students to be self-disciplined both in and out of class. Based on questionnaires, previous studies showed that to improve the regularity of class attendance \cite{Crede2010,Hijazi2006} and to cultivate regular studying habits \cite{Crede2008} will enhance the academic performance. However, these studies have not distinguished orderness in living patterns from diligence in study, since more regular studying habits will result in longer studying time. To our knowledge, a clear and quantitative relationship between orderness in living patterns and academic performance of college students has not yet been unfolded in the literature. Fortunately, with the large-scale behavioral data, especially the extracurricular behavioral records, we are able to quantitatively measure the orderness of a student's lifestyle.

In most Chinese universities, every student owns a campus smart card with real-name registration, which can be used for student identification, and is the unique payment medium for many consumptions in the campus. Therefore, smart cards record large volume of behavioral data in terms of students' living and studying activities, including taking showers in dormitories, having meals in cafeterias, entering and exiting the library and dormitories, fetching water in teaching buildings, and so on. The data considered in this paper includes 25,746,759 consumption records and 3,466,020 entry-exit records, which are collect from 18,960 anonymous undergraduates from September 2009 to March 2015 in UESTC. More details about the data are presented in Data Description of \emph{Materials and Methods}.

According to the above data set, taking a specific behavior, say showers, as an example, if the starting times of showers of student \emph{A} always fall into the range [21:00, 21:30] while student \emph{B} may take a shower at any time, we could say student \emph{A} has a higher orderness than student \emph{B} for showers. Next, we turn to the mathematical issue in quantifying the orderness of a student. Again, considering a specific behavior (\emph{e.g.}, showers, meals, \emph{etc}.) of an arbitrary student, with in total $n$ recorded actions happening at time stamps $\{t_1, t_2, \cdots, t_n\}$, where $t_i \in [00:01, 24:00]$. Notice that, we arrange all actions according to their happening times, namely the $i$th action happens before the $j$th action if $i<j$, while we do not record the dates. We first divide one day into $48$ time bins, each of which spans $30$ minutes and is encoded from $1$ to $48$ (\emph{e.g.}, 0:01-0:30 is the first bin, 0:31-1:00 is the second bin, ...). Then, the time series $\{t_1, t_2, \cdots, t_n\}$ can be mapped into a discrete sequence $\{t'_1, t'_2, \cdots, t'_n\}$ where $t'_i \in \{1,2,\cdots,48\}$. For example, if a student's starting times of five consecutive showers are \{21:05, 21:33, 21:13, 21:48, 21:40\}, the corresponding binned sequence is $\mathcal{E}=\{43, 44, 43, 44, 44\}$. In this paper, we apply the \emph{actual entropy}~\cite{Kontoyiannis1998,Song2010} to measure the orderness of any sequence $\mathcal{E}$, namely
\begin{equation}\label{eq:ae}
	S_{\mathcal{E}}=(\frac{1}{n}\sum_{i=1}^n\Lambda_i)^{-1}\ln{n},
\end{equation}
where $\Lambda_i$ represents the length of the shortest subsequence starting from $t'_i$ of $\mathcal{E}$, which never appeared previously. If such subsequence does not exist, we set $\Lambda_i=0$. Following this definition, given the above sequence, we have $\Lambda_1=1$, $\Lambda_2=1$, $\Lambda_3=3$ and $\Lambda_4=\Lambda_5=0$, and thus $S_{\mathcal{E}}=1.609$. The actual entropy is considered as a metric for orderness: the smaller the entropy, the higher the orderness. The advantages of using actual entropy instead of some other well-known metrics, such as information entropy \cite{Shannon1948} and diversity coefficient \cite{Simpson1949}, are presented in \emph{Appendix A}.

\begin{figure}[ht]
	\centering
	\includegraphics[width=0.48\textwidth]{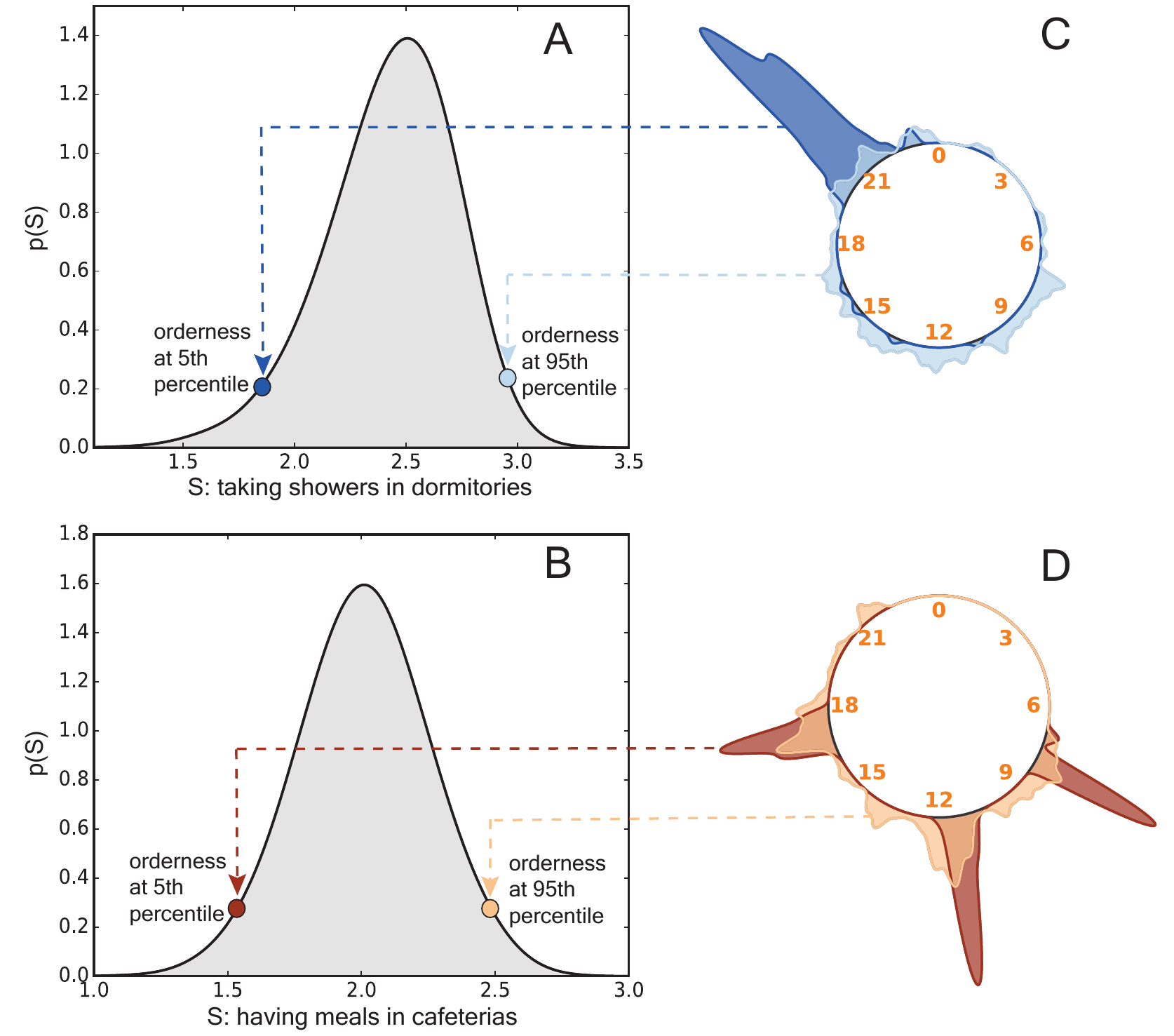}
	\caption{The distributions of actual entropies, $p(S)$, of students in ({\bf A}) taking showers in dormitories and ({\bf B}) having meals in cafeterias. The broad distributions guarantee the discriminations of students with different orderness. To better illustrate the differences in behavioral patterns, the behavioral clocks of two students at the 5th percentile and the 95th percentile are shown for ({\bf C}) taking showers in dormitories and ({\bf D}) having meals in cafeterias. Intuitively, the students with higher orderness have more concentrated behaviors over time whilst the students with lower orderness have much more dispersed temporal activities. The huge differences between their behavioral patterns demonstrate the relevance of the orderness measure. }
    \label{fig2}
\end{figure}

Among various daily activities in campus, we calculate orderness based on two behaviors: taking showers in dormitories and having meals in cafeterias. The reasons to choose these two behaviors are fourfold: (i) they are both high-frequency behaviors so that we have a large number of records; (ii) the data is unobtrusive and thus can objectively reflects the students' lifestyles without any experimental bias; (iii) they are not directly related to diligence; (iv) they are less affected by the specific course schedules since any schedule will leave time for meals and showers. Figure~\ref{fig2}(A) and Figure~\ref{fig2}(B) show the distributions $p(S)$ of actual entropies of students in taking showers in dormitories and having meals in cafeterias, respectively. The broad distributions guarantee the discriminations of students with different orderness. Figure~\ref{fig2}(C) compares two typical students respectively with very high orderness (at the 5th percentile of the distribution $p(S)$, named as student H) and very low orderness (at the 95th percentile of the distribution $p(S)$, named as student L). As clearly shown by the behavioral clock, student H takes most showers around 21:00 while student L may take showers at any time in a day only except for a very short period before dawn, from about 2:30 to about 5:00. Similar discrepancy between two students respectively with very high and very low orderness in having meals can be observed in Fig.~\ref{fig2}(D). In a word, students with higher orderness have more concentrated behaviors over time whilst students with lower orderness have much more dispersed temporal activities.

In addition to orderness, we have also considered another high-level behavioral character called diligence, which estimates the effort a student makes in his/her academic studies. Specifically, we quantify diligence based on two behaviors: entering and exiting the library and fetching water in teaching buildings (see \emph{Appendix B} for details). Empirical analysis also demonstrates that the corresponding distributions are broad enough to distinguish students with different diligence (see Fig. S1 in \emph{Appendix B}).

\subsection*{Analysis}
Intuitively, students with higher orderness are probably more self-disciplined, which is an intrinsic personality not only affecting meals and showers, but also acting on their studying behaviors. Hence we would like to see whether orderness is correlated with academic performance, say GPA. The orderness is simply defined as $O_\mathcal{E}=-S_\mathcal{E}$ and both orderness and GPA are firstly regularized via \emph{Z-score} \cite{Kreyszig2007}, namely
\begin{equation}
    O'_{\mathcal{E}} = \frac{O_{\mathcal{E}}-\mu_O}{\sigma_O} = \frac{\mu_S-S_{\mathcal{E}}}{\sigma_S},
\end{equation}
where $O'_{\mathcal{E}}$ is the regularized orderness for the student with binned sequence $\mathcal{E}$, $\mu_O$ and $\sigma_O$ are the mean and standard deviation of $O$ for all considered students, and $\mu_S$ and $\sigma_S$ are the mean and standard deviation of $S$ for all considered students. Obviously, $\mu_O=-\mu_S$ and $\sigma_O=\sigma_S$. Analogously, the regularized GPA for an arbitrary student $i$ is defined as
\begin{equation}
    G'_i = \frac{G_i-\mu_G}{\sigma_G},
\end{equation}
where $G_i$ is the GPA of student $i$, and $\mu_G$ and $\sigma_G$ are the mean and standard deviation of $G$ for all considered students.

\begin{figure}[ht]
	\centering
	\includegraphics[width=0.40\textwidth]{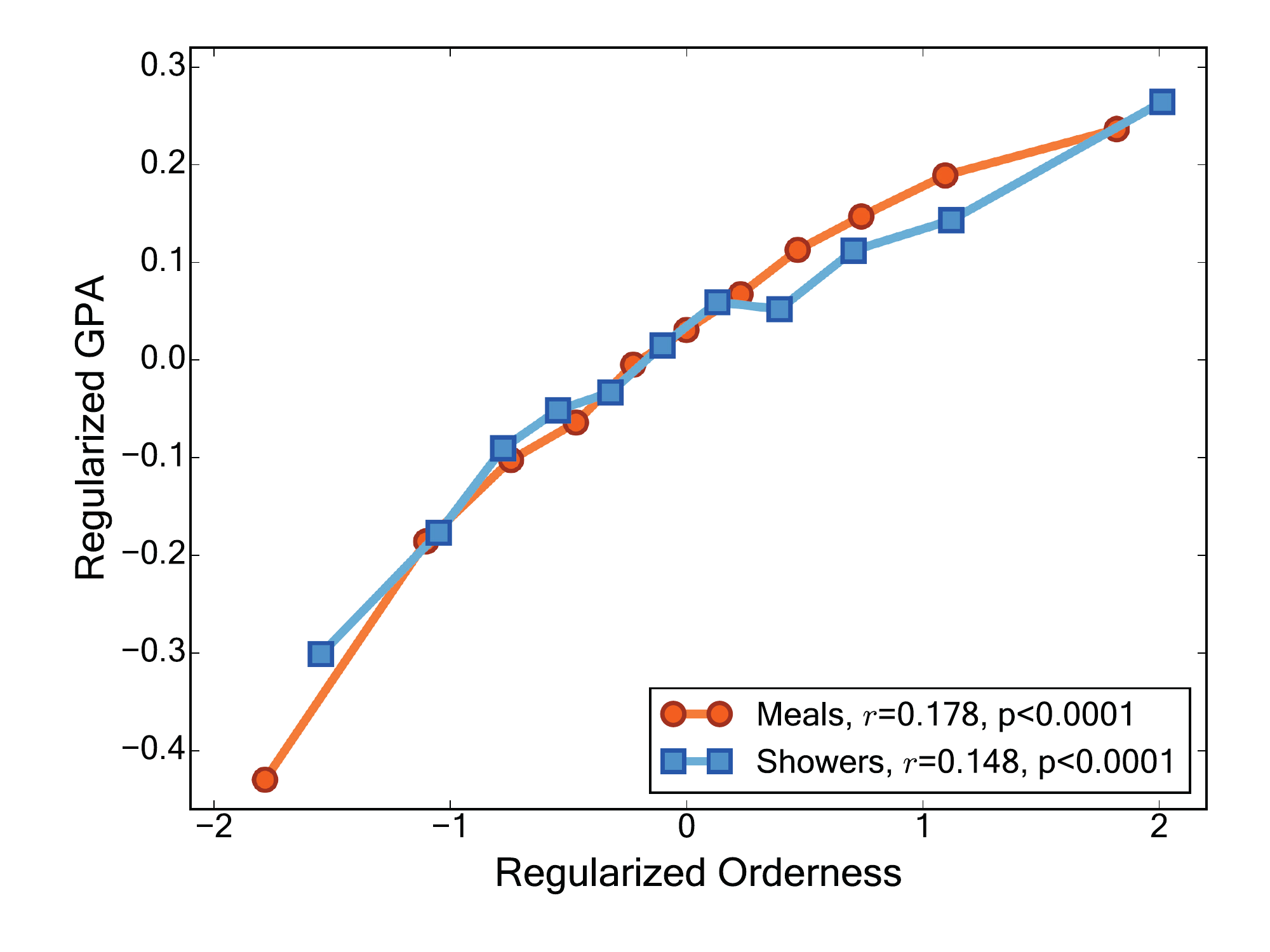}
	\caption{ Relationship between orderness and academic performance for showers (blue squares) and meals (orange circles). The Spearman rank correlation coefficients for GPA-meals and GPA-showers are 0.178 and 0.148 with $p$-values both less than 0.0001. }
    \label{fig3}
\end{figure}

Figure~\ref{fig3} shows the relationships between regularized GPAs and regularized orderness for meals and showers, indicating significantly positive correlations. We apply the well-known Spearman's rank correlation \cite{Spearman1904} to quantify the correlation strength, which is defined as
\begin{equation}
    r_S = 1-\frac{6\sum^N_{i=1} d_i^2}{N(N^2-1)},
\end{equation}
where $N$ is the number of students under consideration, $d_i=r(O'_i)-r(G'_i)$, with $r(O'_i)$ and $r(G'_i)$ being the ranks for student $i$'s orderness and GPA, respectively. The Spearman's correlation coefficient lies in the range $[-1,1]$, and the larger the absolute value is, the high the correlation is. As shown in Fig.~\ref{fig3}, the correlation coefficients for meals and showers are respectively 0.178 and 0.148, with both $p$-values smaller than 0.0001, suggesting the statistical significance.

The significant correlation implies that orderness can be considered as a feature class to predict students' academic performance. Diligence is also significantly correlated with academic performance (see Fig. S2 in \emph{Appendix B}) and thus considered as another feature class in the predicting model. We apply a well-known supervised learning to rank algorithm named RankNet \cite{Burges2005} to predict the ranks of students' semester grades (see the detailed explanation of RankNet algorithm in \emph{Materials and Methods}). We train RankNet based on the extracted orderness and diligence values in one of the first four semesters and predict students' ranks of grades in the next semester. We use AUC value \cite{Hanley1982} to evaluate the prediction accuracy, which, in this case, is equal to the percentage of student pairs whose relative ranks can be consistently predicted with the ground truth. The AUC value ranges from 0 to 1 with 0.5 being the random chance, therefore to which extent the AUC value exceeds 0.5 can be considered as the prediction power. As shown in Table~\ref{tb1}, we calculate AUC values under different feature combinations. It is obvious that both orderness and diligence are effective for predicting academic performance in all testing semesters, and the introduction of orderness can remarkably improve the prediction accuracy even at the presence of diligence. At the same time, we have checked that orderness and diligence are not significantly correlated (see \emph{Appendix C}). That is to say, orderness has its independent effects on academic performance.

\begin{table}[htp]
	\centering
	\caption{AUC values for the GPA prediction. The abbreviations O, D and O+D stand for utilizing features on orderness only, on diligence only and on the combination of orderness and diligence, respectively. SEM is short for semester, for example, SEM 3 represents the case we train the data of semester 2 and predict the ranks of examination performance in semester 3.}
	\begin{tabular}{lrrrr}
		\toprule
		Features & SEM 2 & SEM 3 & SEM 4 & SEM 5\\
		\midrule
		O & 0.618 & 0.617 & 0.611 & 0.597 \\
		D & 0.630 & 0.655 & 0.663 & 0.668 \\
		O+D & 0.668 & 0.681 & 0.685 & 0.683 \\
		\bottomrule
	\end{tabular}
	\label{tb1}
\end{table}

\section*{Discussion}
In this paper, we propose novel metrics to measure two high-level behavioral characters, orderness and diligence, in university campus. Extensive empirical analyses on tens of millions of digital records show strong correlations between orderness and academic performance, as well as between diligence and academic performance. Of particular interests, orderness is calculated from temporal records of taking showers and having meals, which are not directly related to studying behaviors. We further show the considerable predictive power of orderness for academic performance. Compared with most previous works in the literature, this work is characterized by the large-scale unobtrusive data that allows robust statistical analyses.

The majority of known studies in this domain are based on questionnaires with sample sizes usually scaling from dozens to hundreds \cite{Poropat2009,Vedel2014,Crede2010,Crede2008,Dewald2010}. In addition, these studies suffer from experimental bias since subjects would like to report social desirable information while not to report disapproved behaviors \cite{Fisher1993,Paulhus2007}. Therefore, analyzing large-scale unobtrusive digital records will become a promising or even mainstream methodology in the near future. However, we do NOT think such \emph{big-data analyses} should replace questionnaire surveys, instead, these two methodologies will complement and benefit each other. First of all, with the help of large-scale accessible data on individual daily routines, we can estimate the discrimination of a set of items in a questionnaire on the target behavioral character. Therefore, it is very possible that psychologists and computer scientists will work together not only to make use of unobtrusive digital records, but also to improve the quality of questionnaires \cite{Markowetz2014,Montag2016}. Secondly, a few recent works \cite{Chittaranjan2013,Kosinski2013,Youyou2015} show the potential to predict personality and some other private attributes by behavioral data. If such kind of reverse predictions are accurate enough comparing with diagnoses, human judgments, self-reports and questionnaire surveys, then we are able to infer questionnaire results of a large population based on the combination of behavioral records and questionnaires of a small fraction of population.

The present report is highly relevant to education management. On the one hand, understanding the explicit relationship between behavioral patterns and academic performance could help education administrators to guide students to behave like excellent ones and then they may really become excellent later on. In fact, we have shown the average behavioral profile over students with top-5\% GPAs in UESTC, including how much time they spent in the library, how many books and which are the most popular ones they borrowed, how many elective courses and which are the most popular ones they took, and so on. On the other hand, we can detect undesirable abnormal behaviors in time and thus implement effective interventions in an early stage. For example, the Internet addiction is the uppermost reason resulting in the failure of college study in China. There are two lags in the Internet addiction. Firstly, it requires a long time (usually a few months) for a student to rebuild his/her learning ability after getting rid of Internet addiction. Secondly, the sharp fall of examination performance or even the failure of many courses appears about one or two semesters after getting into the Internet addiction. Therefore, to find Internet addicts in an early stage is critical for effective interventions. Being addicted, a student's behavioral pattern is largely different from the ones who are not addicted, in particular, the student's diligence and orderness will be dramatically declined. Some other behavioral patterns not yet being studied in this paper also change a lot if one gets addicted to Internet games, such as the temporal records of entering and exiting dormitories. Accordingly, we will be able to build up models to predict whether a student is an Internet addict and thus implement necessary interventions to avoid invocatable failures in his/her college study.

We believe the reported approaches, together with some other works \cite{Wang2014,Wang2015,Chittaranjan2013,Kosinski2013,Youyou2015} in the same direction, will induce methodological and ideational shifts in pedagogy, eventually resulting in quantitative and personalized education management in the future.

\section*{Materials and Methods}
\subsection*{Data Description}
The data contains four kinds of daily behaviors in UESTC campus. Specifically, there are 3,380,567 records for taking showers in dormitories, 20,060,881 records for having meals in cafeterias, 3,466,020 records for entering and exiting the library, and 2,305,311 records for fetching water in teaching buildings, respectively. For the 18,960 anonymous students under consideration, the data covers from the beginning of their first year to the end of their third year. In addition, some other consumption and entry-exit behaviors are also recorded, including purchasing daily necessities in campus supermarkets, doing the laundries, having coffees in cafes, taking school buses, entering and exiting the dormitories, and so on. GPAs of undergraduate students at each semester are also collected.

\subsection*{Prediction Approach}
Given a characteristic feature vector $\mathbf{x}\in \mathbb{R}^p$ of each student, a pair-wise learning to rank algorithm, RankNet~\cite{Burges2005}, has been exploited to predict students' academic performance. RankNet tries to learn a scoring function $f:\mathbb{R}^p \rightarrow \mathbb{R}$, so that the predicted rankings according to $f$ are as consistent as possible with the ground truth. In RankNet, such consistence is measured by cross entropy between the actual probability and the predicted probability. Based on the scoring function, the predicted probability that a student $i$ has a higher GPA than another student $j$ (denoted as $i\triangleright j$) is defined as $P(i\triangleright j)=\sigma(f(\mathbf{x}_i)-f(\mathbf{x}_j))$, where $\sigma(z)=1/(1+e^{-z})$ is a sigmoid function. Here we consider a simple regression function $f=\mathbf{w}^T\mathbf{x}$, where $\mathbf{w}$ is the vector of parameters. Therefore the cost function of RankNet is formulated as follows:
\begin{equation}
	\mathcal{L}=-\sum_{(i,j):i\triangleright j}\log\sigma(f(\mathbf{x}_i)-f(\mathbf{x}_j)) + \lambda \Omega(f),
\end{equation}
where $\Omega(f)$ is a regularized term to prevent over-fitting, defined as $\Omega(f)=\mathbf{w}^T\mathbf{w}$. Given all students' feature vectors and their rankings, we apply gradient decent to minimize the cost function, where the gradient of the lost function with respect to parameter $\mathbf{w}$ in $f$ is:
\begin{equation}
	\frac{\partial \mathcal{L}}{\partial \mathbf{w}}=\sum_{(i,j):i\triangleright j}(\sigma(f(\mathbf{x}_i)-f(\mathbf{x}_j)) - 1) (\frac{\partial f(\mathbf{x}_i)}{\partial \mathbf{w}} - \frac{\partial f(\mathbf{x}_j)}{\partial \mathbf{w}}) + \lambda \frac{\partial\Omega(f)}{\partial\mathbf{w}}.
\end{equation}

\section*{Acknowledgments}
The authors would like to acknowledge Hao Chen, Yan Wang in Nankai University, Qin Zhang, Junming Huang and Jiansu Pu in UESTC for valuable discussions. This work was partially supported by the National Natural Science Foundation of China (61603074, 61473060, 61433014, 61502083).

\section*{Appendix A: Actual entropy and other measures on orderness}
The meaning of \emph{orderness} is twofold, say timing and order. Firstly, the happening times of the same kind of events should be close to each other, for example, taking breakfast at around 8:00 in the morning is more regular than taking breakfast between 7:00 and 9:00. Secondly, the temporal order of the occurrences of different kinds of events should be regular. For instance, one may go to the cafeterias following the order: breakfast$\rightarrow$lunch$\rightarrow$supper$\rightarrow$breakfast$\rightarrow$lunch$\rightarrow$supper, which is more regular than breakfast$\rightarrow$supper$\rightarrow$lunch$\rightarrow$ supper$\rightarrow$breakfast$\rightarrow$lunch.

We apply actual entropy \cite{Kontoyiannis1998}, since it takes into account both the above two ingredients. It is defined as:
\begin{equation}\label{eq:a_ae}
	S_{\mathcal{E}}=(\frac{1}{n}\sum_{i=1}^n\Lambda_i)^{-1}\ln{n},
\end{equation}
where n is the length of the sequence, and $\Lambda_i$ represents the length of the shortest subsequence starting from $i$th position of the binned event sequence $\mathcal{E}$, which never appeared previously. If any sequence starting from $i$th position has appeared previously, we set $\Lambda_i=0$. The smaller $S_{\mathcal{E}}$ suggests higher orderness. Considering two consecutive sequences on taking showers with different happening times, which are \{21:05, 21:13, 21:17, 21:28, 21:24, 21:15, 21:12, 21:08, 21:19, 21:03\} and \{21:05, 21:33, 21:13, 21:48, 21:40, 21:15, 21:42, 21:18, 21:49, 21:53\}, respectively. Accordingly, the two binned sequences are \{43, 43, 43, 43, 43, 43, 43, 43, 43, 43\} and \{43, 44, 43, 44, 44, 43, 44, 43, 44, 44\}, and the corresponding length sequences $\Lambda$ are \{1, 2, 3, 4, 5, 0, 0, 0, 0, 0\} and \{1, 1, 3, 2, 4, 0, 0, 0, 0, 0\}. According to Eq.~\ref{eq:a_ae}, the actual entropy values are 1.535 and 2.093, suggesting that the former sequence is more regular than the latter. Next, we take meals as an example. Assume that both two students under consideration take breakfast, lunch and supper at 7:30-8:00, 11:30-12:00 and 17:30-18:00, but one has every meal in the campus, while the other sometimes has a dinner party outside or does not take breakfast. The binned sequences of the two students could be \{16, 24, 36, 16, 24, 36, 16, 24, 36, 16, 24, 36, 16, 24, 36\} and \{16, 36, 16, 16, 24, 36, 36, 16, 24, 36, 16, 24, 24, 24, 36\}. The corresponding length sequences $\Lambda$ are \{1, 1, 1, 4, 4, 4, 7, 7, 7, 0, 0, 0, 0, 0, 0\} and \{1, 1, 2, 2, 1, 2, 3, 4, 3, 4, 3, 2, 3, 0, 0\}, and the actual entropy values are 1.128 and 1.310, respectively. The result illustrates that the former sequence is more regular than the latter, and if the sequences get longer, the differences will be more remarkable.

\begin{figure}[ht]
    \centering
    \includegraphics[width=0.48\textwidth]{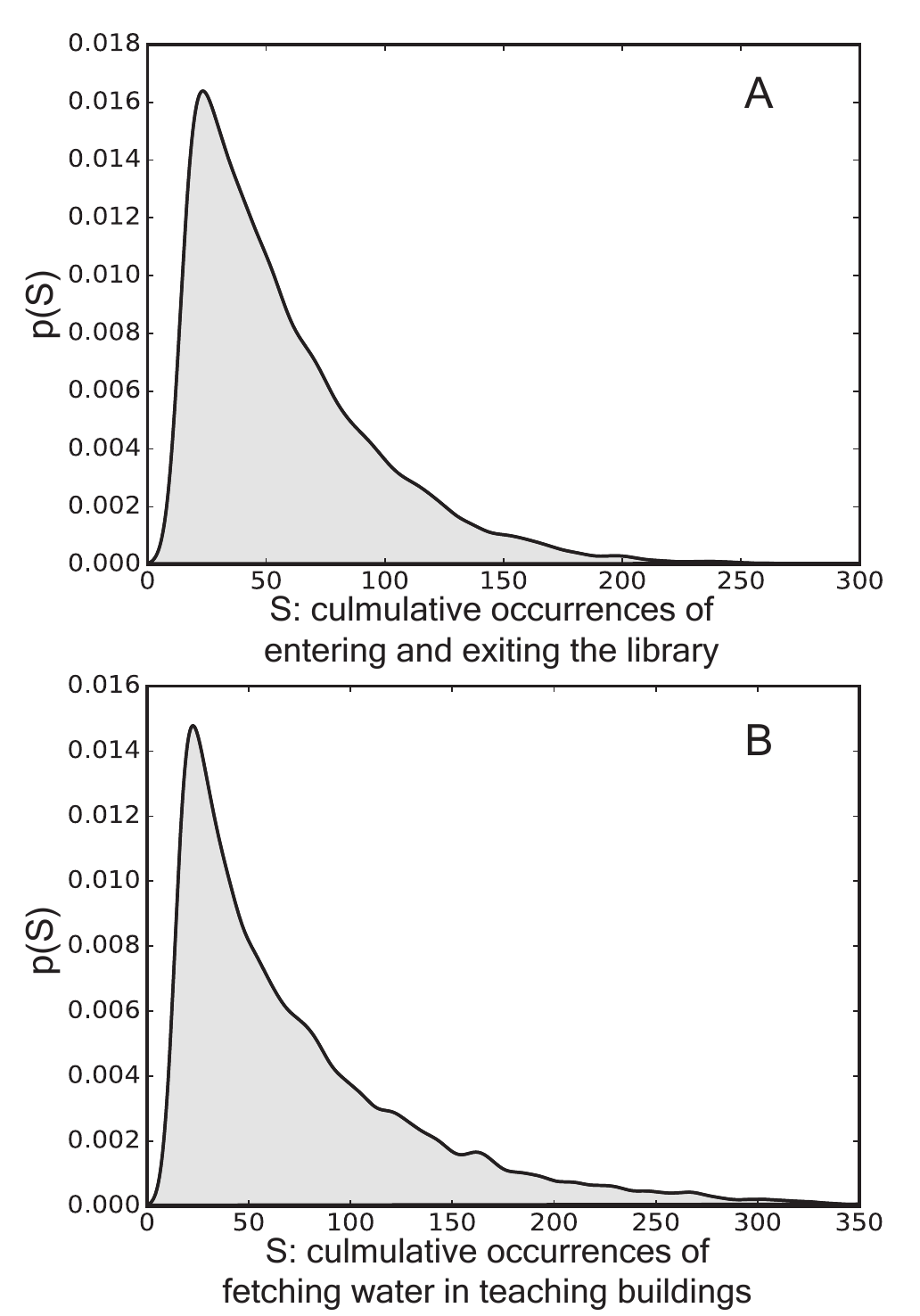}
	\caption{The distributions of two diligence metrics, $p(S)$, of students for ({\bf A}) entering and exiting the library and ({\bf B}) fetching water in teaching buildings. The broad distributions of cumulative occurrences ensure that students with different diligence levels are distinguishable from each other.}
    \label{fig.s1}
\end{figure}

Besides actual entropy, we come up with a few classic metrics to explain why the temporal order of a behavioral sequence is so important and why these metrics are inappropriate to quantify the orderness. Information entropy \cite{Shannon1948} is the most frequently used metric to measure the regularity, which is defined as
\begin{equation}\label{eq:a_infoE}
E=-\sum_{i=1}^{N}p_i\log_2{p_i},
\end{equation}
where $N$ is the number of different kinds of elements (here $N=48$), and $p_i$ denotes the normalized frequency of the $i$th element, and thus
\begin{equation}\label{eq:a_prob}
\sum_{i=1}^{N}p_i=1.
\end{equation}
Larger $E$ means higher orderness. However, this method cannot distinguish sequences of happening times with different temporal order. For example, the two students with different meal times, as shown above, are assigned exactly the same information entropy, since the normalized appearance frequencies of 16, 24 and 36 are all 1/3.

\begin{figure}[ht]
    \centering
    \includegraphics[width=0.48\textwidth]{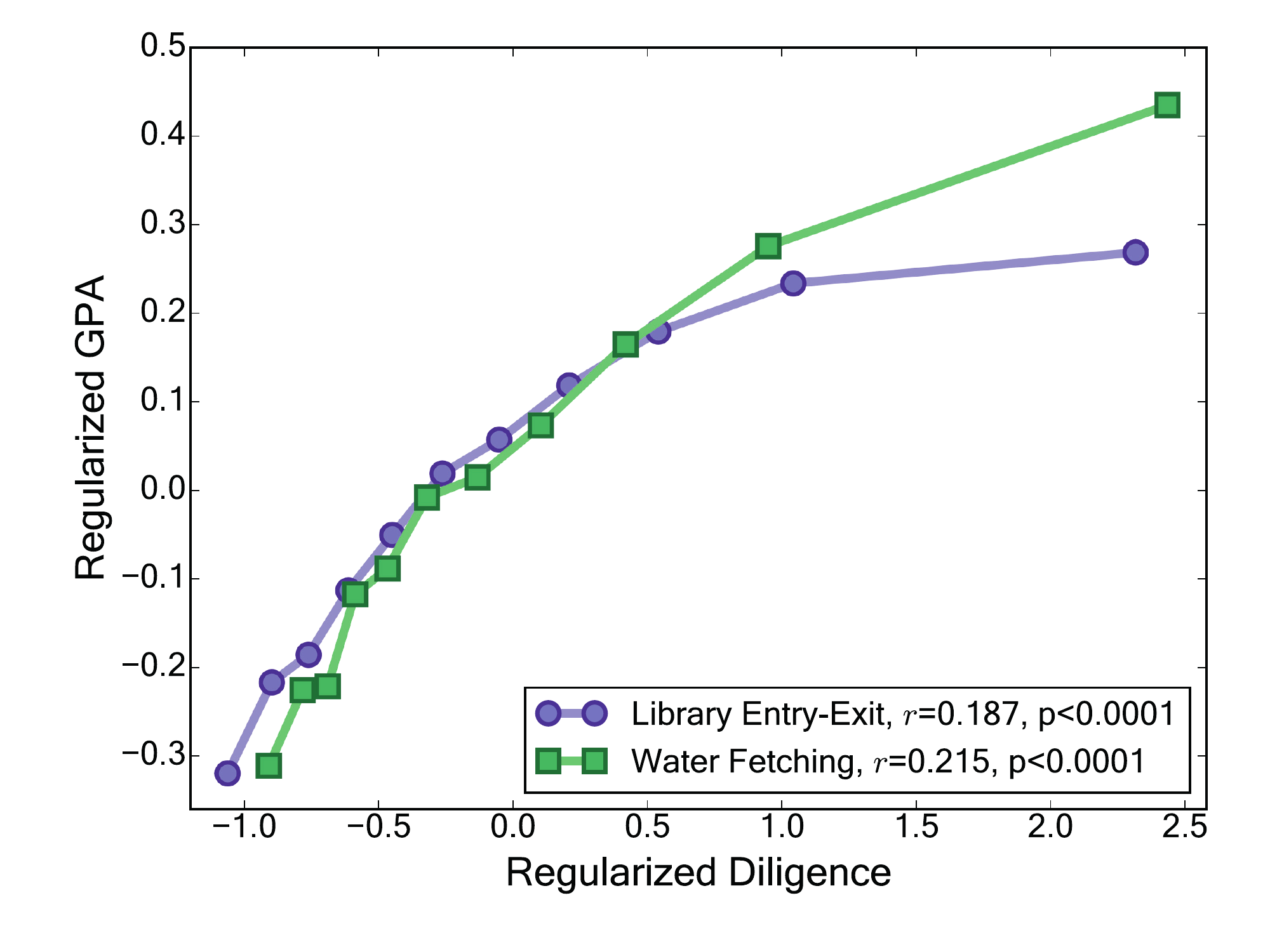}
    \caption{Relationship between diligence and academic performance for entering and exiting the library (purple circles) and fetching water in teaching buildings (green squares). The Spearman correlation coefficients for GPA-library and GPA-water are 0.187 and 0.215 with $p$-values both less than 0.0001.}
    \label{fig.s2}
\end{figure}

Analogously, the well-known Simpson index \cite{Simpson1949} also cannot differentiate the differences when measuring the temporal order. The Simpson index is initially used to measure the diversity of entities when they are classified into different types. Here, it has been extended to represent the regularity level of a given behavioral sequence. It is defined as:
\begin{equation}\label{eq:a_simpson}
D=\frac{\sum_{i=1}^{N}r_i(r_i-1)}{n(n-1)},
\end{equation}
where $r_i$ is the number of appearances of the $i$th element, say
\begin{equation}\label{eq:a_freq}
\sum_{i=1}^{N}r_i=n.
\end{equation}
A sequence has higher orderness if $D$ is larger. Considering the above two students with different meal times, based on Eq.~\ref{eq:a_simpson}, the Simpson index values are the same, say $D=0.286$.

In a word, the above two classic metrics are inappropriate to measure the orderness since they only consider the number of the events but do not care about the temporal order of these events.

\section*{Appendix B: Diligence}
Diligence is another high-level behavioral character that stands for how people control themselves to strive for achievements, and thus it is considered as a class of high-level features that is directly correlated with academic performance. In this paper, two behaviors are used to depict and evaluate the diligence: entering and exiting the library, and fetching water in teaching buildings. Normally, borrowing books and self-studying are the most common purposes of a student to go to the library, while attending professional courses is the most common purpose of being at teaching buildings. However, unlike the library, the teaching buildings have no entry terminals or check-in devices. Hence we select fetching water as a proxy behavior with high frequency. For each behavior, we use the cumulative occurrences to estimate the diligence level. We show the distributions of diligence metrics in Fig.~\ref{fig.s1}. The two diligence metrics are both broadly distributed, suggesting that the two metrics are good to distinguish students with different levels of striving for achievements. We show the correlation between diligence and GPA in Fig.~\ref{fig.s2}, where both metrics and GPA are regularized by Z-score \cite{Kreyszig2007}. We apply the Spearman's rank correlation coefficient\cite{Spearman1904} to quantify the correlation between regularized GPAs and regularized diligence metrics, and show the result in Fig.~\ref{fig.s2}. Obviously, academic performance is vitally and positively correlated with diligence for both entering and exiting the library and fetching water in teaching buildings. The corresponding Spearman correlation coefficients are respectively 0.187 and 0.215, with both $p$-values less than 0.0001, indicating the significance of the correlation.

\section*{Appendix C: Correlation between orderness and diligence}
\begin{figure*}[htp]
    \centering
    \includegraphics[width=0.65\textwidth]{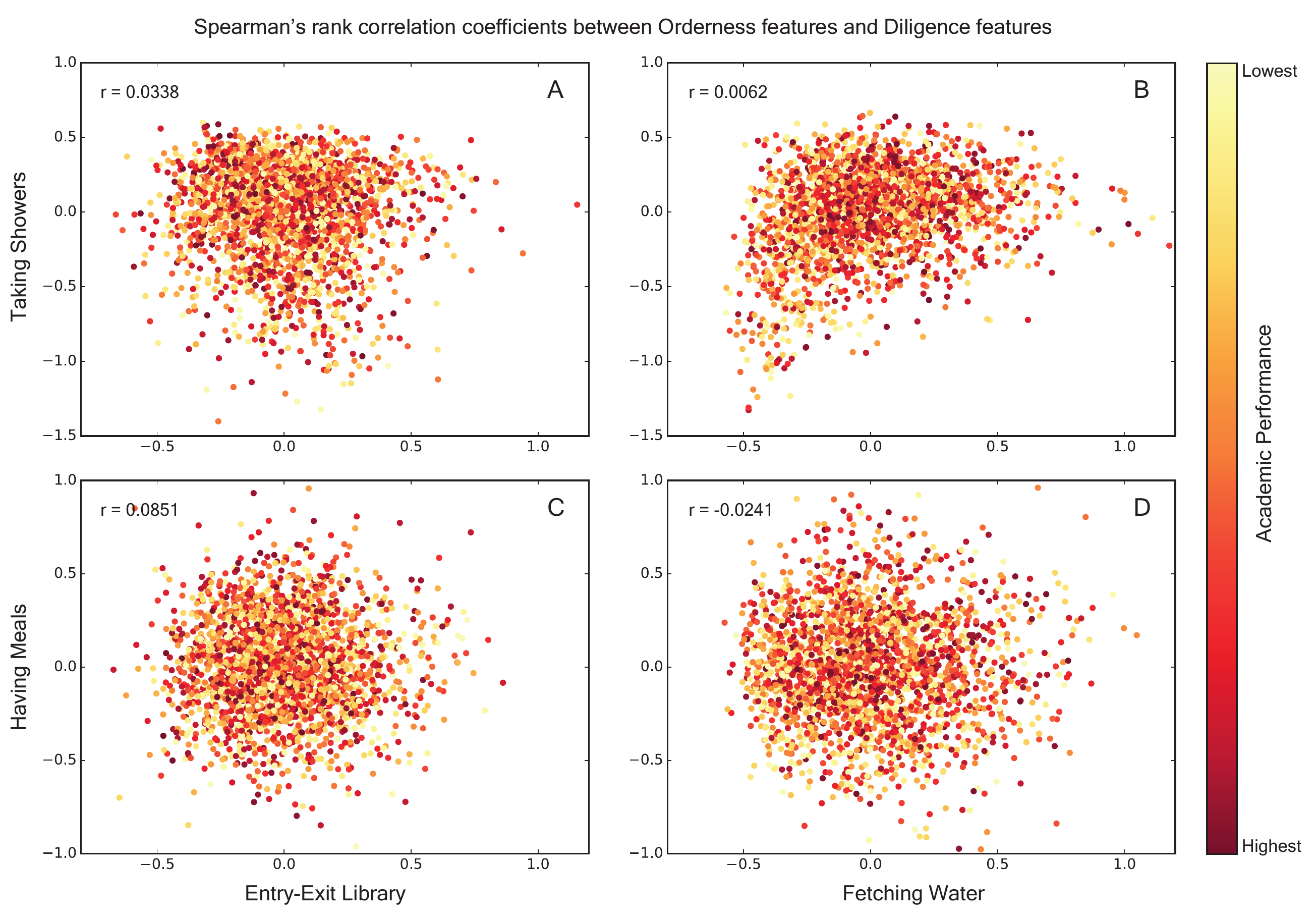}
    \caption{Relationship between orderness and diligence. The colorbar represents the academic performance: the deeper the color, the better the performance.}
    \label{fig.s3}
\end{figure*}

We show the Spearman's rank correlation coefficients between orderness features and diligence features in Table~\ref{tb.s1}. The results indicate that there is no significant correlation between orderness and diligence. To visually illustrate the independence of orderness on diligence, we draw the scatter diagram for the four orderness-diligence feature pairs in Fig.~\ref{fig.s3}, where each data point represents a student, and the better-performed students are assigned with deeper color. As shown in Fig.~\ref{fig.s3}, there has no multicollinearity problem between orderness and diligence.

\begin{table}[htp]
	\centering
	\caption{The Spearman's rank correlation coefficients between orderness features and diligence features.}
	\begin{tabular}{lrr}
		\toprule
		 & Entry-Exit Library & Fetching Water \\
		\midrule
		Taking Showers & 0.0338 & 0.0062 \\
		Having Meals & 0.0851 & -0.0214 \\
		\bottomrule
	\end{tabular}
	\label{tb.s1}
\end{table}

In each plot, we show 2500 data points randomly selected from the total 18960 samples. As shown in this figure, the orderness features and diligence features are not dependent to each other for neither the entire population nor a subset of students with very good, middle or poor performance.


\end{document}